# Exploring unconventional superconductivity in PdTe via Point Contact Spectroscopy

Pritam Das[*1†], Sulagna Dutta[*1], Saurav Suman[1], Amit Vashist[2], Bibek Ranjan Satapathy[2], John Jesudasan[1], Suvankar Chakraverty[2], Rajdeep Sensarma[1] and Pratap Raychaudhuri[1‡]

[1]*Tata Institute of Fundamental Research, Homi Bhabha Road, Mumbai 400005.*

[2]*Institute of Nano Science and Technology, Sahibzada Ajit Singh Nagar, Punjab 140306.*

Palladium Telluride (PdTe), a non-layered intermetallic crystalline compound, has captured attention for its unique superconducting properties and strong spin-orbit coupling. In this work, we investigate the superconducting state of PdTe using point-contact Andreev reflection (PCAR) spectroscopy. The experimental data are analysed using the Blonder-Tinkham-Klapwijk (BTK) model for *s*, *p* and *d* wave symmetries. Our results reveal clear evidence of unconventional superconductivity. The superconducting gap showing features consistent with either *p*-wave or *d*-wave pairing symmetries but cannot be fitted with *s*-wave symmetry. The observed anisotropic gap structure and deviations from conventional BCS behaviour highlight the complex nature of the pairing interactions in PdTe. These findings provide strong evidence of unconventional pairing symmetry in this material.

---

[*] The authors have contributed equally.
[†] Email: pritam.das@tifr.res.in
[‡] Email: pratap@tifr.res.in

**Introduction**

Understanding the symmetry of the superconducting order parameter is essential for uncovering the pairing mechanism in superconductors. Unlike conventional *s*-wave superconductors, which have an isotropic energy gap, many newly discovered superconductors exhibit highly anisotropic pairing. This includes high-$T_c$ cuprates [1], the triplet superconductor like $Sr_2RuO_4$ [2], and some heavy fermion compounds [3]. A definitive confirmation of the gap symmetry is crucial since it is directly linked to the pairing mechanism. For example, isotropic electron-phonon attraction favours zero-angular-momentum pairing with a spherically symmetric order parameter. In contrast, *d*-wave symmetry is more likely to emerge when the pairing interaction is short-range repulsive and anisotropic at larger distances, such as the effective spin-spin interactions near antiferromagnetic order [4,5,6]. Similarly, *p*-wave symmetry, which is associated with spin-triplet pairing, often arises in systems with strong spin-orbit coupling or ferromagnetic fluctuations [7,8].

Point-contact Andreev reflection (PCAR) spectroscopy has emerged as a powerful tool to probe the electronic properties of variety of superconductors. This technique directly measures the energy-dependent density of states and provides insights into the superconducting gap structure and underlying quasiparticle interactions [9,10]. It allows for the extraction of energy, momentum, and spin-resolved information regarding the Fermi surface [11,12,13,14]. Andreev reflection [15] is a process, by which an electron is incident on a normal metal-superconductor interface having energy less than the superconducting gap, gets reflected as a hole in the opposite spin band of the normal metal and transmits a cooper pair in the superconductor. A very fine tip made of normal metal is brought into mechanical contact with a superconductor and the differential conductance vs. voltage across the contact is analysed to extract information on the magnitude and symmetry of the superconducting energy gap. This technique has been applied to systems like heavy Fermion superconductors [16,17,18], $BiS_2$ based superconductors [19,20], Fe-based superconductors [21,22], multigap superconductors [11] and layered perovskite without copper [23]. With this technique, one can also measure the degree of spin-polarization in ferromagnetic materials [24,25,26] and topological insulators [27]. It can also effectively probe and extract characteristic of unconventional superconductors including *p*-wave [28,29] and *d*-wave pairing [4].

Palladium Telluride (PdTe) is a binary intermetallic compound that has recently attracted attention due to its unique physical and electronic properties. Unlike transition metal dichalcogenides (TMDs), which are layered, and exhibit van der Waals bonding between planes, PdTe is a non-layered material with a distinct crystallographic structure [30,31,32,33,34,35,36,37,38]. This compound has garnered significant scientific interest due to its striking similarities to the unconventional superconductivity observed in iron-based chalcogenides and pnictides [21,39,40,41].

PdTe is notable for its superconducting properties, with reported critical temperature ($T_c$) varying between 3.8 K to 4.6 K [30,31,32,33,34,35,36,37,38]. Currently, the nature of superconductivity in PdTe is under considerable debate. Various possibilities, such as multiband superconductivity [30,31,32] as well as unconventional pairing symmetries [34] have been proposed in this compound. Recent ARPES [37] studies suggest that its superconductivity exhibits unconventional behaviour of possessing nodal bulk gaps. It has been suggested that this the unconventional pairing could result from an interplay of strong spin-orbit coupling and the electronic band structure [35]. While some reports based on specific heat and resistivity measurement conclude that PdTe is a multiband superconductor [30,31,32], one report [34] down to 50 mK observes a power-law dependence of specific heat with temperature, consistent with unconventional superconductivity. In principle, multiband superconductivity could also co-exist with unconventional pairing symmetry as reported earlier in some compounds [40,41]. It is therefore important to probe the gap symmetry using different spectroscopic probes.

In this paper, we investigate the electronic properties of PdTe using point contact Andreev reflection (PCAR), aiming to elucidate its superconducting electrons pairing and superconducting gap structure. Our data strongly supports unconventional pairing symmetry in this compound.

**Experimental Methods**

High-quality PdTe single crystals were grown using the melt growth technique where high-purity Pd (99.99%) and Te (99.999%) were mixed in a 1:1 molar ratio. (For details about sample growth, see Ref [30]). A Laue pattern obtained using molybdenum white x-ray source (0.5 - 3 Å). Resistance versus temperature measurements performed in 4-probe geometry in a continuous flow cryostat down to 2.3 K. To confirm the bulk nature of superconductivity, ac susceptibility measurement was also performed on another crystal using a two-coil mutual inductance technique [42] with an ac excitation of 0.5 mA ( which gives rise to a peak magnetic field of 8 mOe ) at 33 kHz.

Point contact Andreev reflection (PCAR) measurements were performed on PdTe crystals which were cut into small pieces parallel to the *a-c* plane. Before conducting the point contact measurement, the crystal surface was polished to a mirror-like finish. Our PCAR arrangement [43] is a Needle anvil type point contact set up, where an Ag or Cu wire (0.25 mm in diameter) was mechanically cut by scissor and brought into contact with the crystal at low temperature using a nanopositioner arrangement. Often, we needed to puncture a surface oxide barrier in order to achieve a stable contact. Thus, after a stable contact is achieved, the resistance of the contact is fine-tuned by retracting the tip from the sample in small steps to achieve a contact for which the contact diameter is smaller than the electronic mean free path. The conductance spectra were then obtained by numerically differentiating the *I-V* data acquired in a 4-probe configuration.

**Results and Discussion**

Fig. 1. (a) shows the crystallographic structure of PdTe. It has a NiAs-type hexagonal crystal structure with P6$_3$/mmc (194) space group and has lattice parameters a = b = 4.152 Å and c = 5.671 Å [30,31,32,33,34,35,36,37,38] PdTe cannot be exfoliated using scotch tape, so it is not a layered material. The Laue pattern is showed in Fig. 1. (b). The experimentally observed Laue diffraction pattern closely matched the simulated pattern obtained using Orient Express V3.3 software for the incident x-ray incident on the *a-c* plane of the crystal [44]. This similarity suggests that the sample has good crystallinity. Fig. 1. (c) shows the temperature-dependent resistivity ($\rho$ vs T) of a single crystal of PdTe. The resistance decreases with temperature, showing typical metallic behaviour with a superconducting transition temperature T$_c$ ~ 4.2 K. The residual resistivity ratio, RRR = $\rho$(T = 280K)/$\rho$(T = 4.2K) ≈ 32 is similar to the value reported in ref. [38] and much larger than that reported in ref. [34]. Fig. 1 (d) shows the ac susceptibility measured on another crystal with T$_c$ ~ 3.9 K. The crystal shows clear diamagnetic response in $M'$ below 3.9 K and a single dip in $M''$ close to T$_c$ (corresponding to the dissipative peak) confirming the bulk nature of superconductivity.

Fig. 2. shows normalised PCAR spectra for 6 different contacts (C1-C6) between the sample and the tip taken at the lowest temperatures. In each spectrum the value of the conductance is normalised to a value larger than that superconducting gap. The contacts were always nominally established on the *a-c* plane. However, due to the roughness of the mechanically polished surface, there is always considerable uncertainty in the direction. To get reliable spectroscopic information we need to ensure that the contact is in the ballistic or diffusive regime such that the energy information of the electron is not lost while passing through the point contact [45]. For this the contact diameter, *D*, has to be smaller than the electronic mean free path, *l*. When $D > l$ the point contact resistance is dominated by inelastic scattering in the contact. It has been shown that in this regime and the point contact $\frac{dI}{dV}$ vs $V$ can exhibit unusual features associated with the critical current of the superconductor that could be mistaken for unconventional superconductivity [46,47,48]. As a first check, we estimate the diameter (*D*) of the contact using the Wexler formula [49,50,51], $R_C = \frac{16\rho l}{3\pi D^2} + \frac{\rho}{D}$, where the first term arises from Sharvin resistance and the second term arise from Maxwell resistance. Here $\rho$ is resistivity, *l* is electronic mean free path and $R_C$ is the contact resistance. For our sample $\rho(4.2K)$ ~ 1.41 µΩ − cm and *l* ~ 306 nm earlier estimated from crystals with RRR similar to our samples [38]. The contact resistance ($R_C$) of the junctions used in this study varied between 0.17 Ω and 0.34 Ω. This corresponds to *d* ~ 169 to 253 nm, which is smaller than *l*. To further verify that the contacts carry genuine spectroscopic information, for several contacts we investigated the evolution of the PCAR conductance spectra for various contact resistances by gradually retracting the tip from the sample in small steps. Fig. 3(a) and 3(b) show representative spectra recorded over a range of point-contact resistances ($R_C$), from 0.27 Ω to 0.76 Ω and 0.15 Ω to 1.56 Ω, respectively. The corresponding contact diameters vary from 108 nm to 193 nm in (a), and from 73 nm to 273 nm in (b). These diameters are smaller than the electronic mean free path

(*l*), confirming no loss in spectroscopic information [45]. The qualitative features of the spectra do not change except for a change in height of the zero bias conductance peak. Furthermore, the extracted value of the superconducting gap in each set of curves remains the same for all spectra (see Appendix 2). Together, these two observations strongly suggest that the PCAR spectra indeed correspond to intrinsic spectroscopic features of the superconductor. It is only for very low resistance contact such as the ones shown in Fig. 3 (c) we observe dip-like feature that arise from the current through the contact reaching the critical current of the superconductor [46,47,48]. The conductance curves from this kind of low resistance contacts do not contain any spectroscopic information and are excluded from the study.

Next, we try to fit the PCAR spectra in Fig. 2 (b)-(g). First, we try to fit the data assuming isotropic *s*-wave order parameter for the superconductor using the Blonder-Tinkham-Klapwijk (BTK) formalism. Within the BTK model the transmittivity of the contact is characterised by two parameters: the superconducting energy gap, $\Delta$, and a barrier parameter $Z = \frac{2V_0}{\hbar v_F}$ ($v_F$ is the Fermi velocity) that accounts for a potential barrier at the N-S interface (at $z = 0$) that is modelled by a delta-function potential $V = V_0 \delta(z)$. In addition to any physical barrier at the interface, Z also accounts for the effect of Fermi velocity mismatch between the two metals [52,53]. In addition, we also add a broadening parameter $\Gamma$, which accounts for nonthermal sources of broadening [54]. The current through the point contact can be expressed in terms of the Andreev reflection and the normal reflection probabilities, $A(E, \Delta, \Gamma, Z)$ and $B(E, \Delta, \Gamma, Z)$:

$$I_{N/S}(V) \propto v_F \int_{-\infty}^{+\infty} [f(E - eV, T) - f(E, T)][1 + A(E, \Delta, \Gamma, Z) - B(E, \Delta, \Gamma, Z)]dE \quad (1)$$

where *E* is the energy of the incoming quasiparticle with respect to the Fermi energy ($\varepsilon_F$) and *f(E,T)* is the Fermi function. $A(E, \Delta, \Gamma, Z)$ and $B(E, \Delta, \Gamma, Z)$ can be calculated by solving the Bogoliubov–de Gennes (BdG) equations. The explicit expressions are given in ref. [53] (also see Appendix 1). Out of the six spectra shown in Fig. 2 we could reasonably fit only one spectrum with this model with, giving $\Delta$ = 0.58 *meV*. Rest of the spectra showed large deviation from the fitted curves and the qualitative shape could be captured only with unusually low gap values ( $\Delta \approx 0.10~meV$ ) (Table 1). Attempts to extend the fits with 2D BTK formalism with *s*-wave order parameter results in fits that are similar to the ones shown above. This suggests that the gap structure is likely more complex than the isotropic *s*-wave symmetry.

Since *s*-wave order for the superconductor failed to give a satisfactory fit to the PCAR spectra we try to fit the data with unconventional order parameter symmetries. We attempt the fit with two different gap functions: (i) the *d*-wave symmetry $d_{x^2-y^2}$ for which the gap function is given by $\Delta(\boldsymbol{k}) = \widetilde{\Delta} \left( \frac{(k_x^2 - k_y^2)}{k_F^2} \right)$ and (ii) the *p*-wave symmetry $p_y$ for which the gap function is given by, $\Delta(\boldsymbol{k}) = \widetilde{\Delta} \left( \frac{k_y}{k_F} \right)$. To fit the data we use an extension of the BTK formalism proposed by Tanaka [55] For such unconventional gap functions the Andreev reflection process depends strongly on the angle between the direction of the current injection and with respect to the normal vector of the interface, $\theta$, and the orientation $\hat{k}_x$ with respect to the normal vector ($\hat{n}$) of the interface, $\alpha$ (see Fig. 4). Consequently, reflection probabilities $A$ and $B$ become function of $\theta$ and $\alpha$. The complete expressions for $A(E, \widetilde{\Delta}, \Gamma, Z, \theta, \alpha)$ and $B(E, \widetilde{\Delta}, \Gamma, Z, \theta, \alpha)$ for $d$ and $p$ waves are given in the Appendix 1. Furthermore, to account for the angular spread in the direction of current one needs to integrate over $\theta$. The current through the point contact in Eqn. (2) is given by,

$$I_{N/S}(V, \alpha) \propto v_F \int_{-\pi/2}^{\pi/2} \cos\theta \, d\theta \int_{-\infty}^{+\infty} [f(E - eV, T) - f(E, T)] [1 + A(E, \widetilde{\Delta}, \Gamma, Z, \theta, \alpha) - B(E, \widetilde{\Delta}, \Gamma, Z, \theta, \alpha)] dE .$$ (2)

The fit of the spectra with $d$-wave and $p$-wave are shown in Fig. 2. The fits parameters for the two symmetries are given in Table 1. We observe that both symmetries fit the spectra equally well for both the fits with $p$-wave and $d$-wave. More importantly, the extracted values of $\widetilde{\Delta}$ are closely bunched around $1 \pm 0.2\ meV$. The major difference between the spectra whose shapes vary widely, is the parameter $\alpha$, which depends on the orientation of the plane of the N/S interface with respect to crystallographic axes of the crystal. Since the mechanically polished crystal surface is still very rough at microscopic length scales, the local facet on which the point contact is established varies from contact to contact, thereby explaining this variation. Based on the extracted values of $\widetilde{\Delta}$, the $\widetilde{\Delta}/k_B T_C \sim 2.32$ to $2.83$ for the fit with $d$-wave symmetry and $2.66$ to $2.83$ for the fit with $p$-wave symmetry.

Fig. 4 shows the temperature evolution of the normalised PCAR spectra for 3 different contacts (C1, C2 and C6) fitted with $p$-wave ((a)-(c)) and with $d$-wave ((d)-(f)) symmetries respectively [56]. As the sample approaches the critical temperature ($T_c$), the Andreev reflection signal gradually diminishes, eventually flattening out near $T_c$. All spectra can again be fitted equally well with both symmetries and the temperature dependence $\widetilde{\Delta}$ (shown in Fig. 5(a)) are close to each other. Variation of the other fit parameters are shown in Fig. 5(b)-(d). Z and α undergo small variation over the entire temperature range. In principle, these values should remain constant; the small variation very likely arises from small changes in contact orientation with temperature. On the other hand, Γ undergoes irregular variations with temperature that vary from contact to contact. The plausible reason for this is the anisotropic nature of the superconducting gap. For an isotropic $s$-wave superconductor, Γ mostly arises from the recombination of electron-like and hole-like quasiparticles and monotonically increases with temperature [54]. On the other hand, when the gap function is anisotropic, the current in each direction samples a distribution of gap values, which depends on the projection of the Fermi surface in a plane perpendicular to the current [57]. This averaging over the Fermi surface gives an additional contribution to Γ. Thus, when the current injection is predominantly along the node, small variations of α with change in temperature, results in large variation of the effective gap distribution and concomitantly of Γ. We believe that this is main origin for the large non-systematic variation in Γ with temperature.

Fig. 6. (a) and (b) show normalized conductance spectra ($G_N(V)$ $vs$ $V$) (normalised with respect to the conductance value at large bias) with different magnetic fields (H) applied perpendicular to the junction at T = 2.3 K fitted with $p$ and $d$ wave superconducting gap symmetries respectively. The peak remains present at magnetic fields till 850 Oe. At H = 950 Oe the superconducting gap totally vanishes, showing a flat spectrum. Fig. 6 (c) shows the variation of superconducting gap with magnetic fields and going to zero at a field value of 950 Oe at T = 2.3 K. From this magnetic field variation, we obtain the H$_{c2}$(0) ≈ 1890 Oe by using $H_{c2} = H_{c2}(0)[\frac{(1-(\frac{T}{T_c})^2)^2}{(1+(\frac{T}{T_c})^2)^2}]$ from G-L theory. The value matches with previous reports [30,31,38].

**Conclusion**

To summarise, we utilized point-contact Andreev reflection spectroscopy (PCAR) to probe the superconducting energy gap and the pairing symmetry of the gap function in PdTe. Our study unambiguously rules out the possibility of $s$-wave symmetric superconducting energy gap in PdTe and strongly supports the existence of nodes in the gap function. The PCAR spectra fit well with both $p$ and $d$ wave. This represents the limitation of this technique in determining the precise gap symmetry from the gap function from the shape of the spectra alone and call for further measurements such as SQUID interference measurements [1] and muon spin rotation to pinpoint the exact symmetry of the gap function. Our results are consistent with ARPES and specific heat measurements which also indicate unconventional superconductivity and the presence of nodes in the gap function. In principle, scanning tunnelling spectroscopy (STS) can provide further insight on the nature of the gap function. However, since the crystals proved difficult to cleave to expose an atomically smooth facet, it would be interesting to attempt growth of PdTe thin films on which STS measurements can be carried out.

**Appendix 1**

For $s$-wave we have used the one-dimensional BTK model. Here the reflection and transmission coefficients $A(E,\Delta,\Gamma,Z)$ and $B(E,\Delta,\Gamma,Z)$, for a nonmagnetic metal / $s$-wave isotropic superconductor interface are represented by $A(E,\Delta,\Gamma,Z) = |a(E,\Delta,\Gamma,Z)|^2$ and $B(E,\Delta,\Gamma,Z) = |b(E,\Delta,\Gamma,Z)|^2$ shown in Eqn. (3), where,

$$a(E,\Delta,\Gamma,Z) = \frac{4\Delta}{(4+Z^2)((E+i\Gamma)+\Omega)-Z^2((E+i\Gamma)-\Omega)} \text{ and } b(E,\Delta,\Gamma,Z) = \frac{-2Z(Z+2i)\Omega}{(4+Z^2)((E+i\Gamma)+\Omega)-Z^2((E+i\Gamma)-\Omega)}$$

$$\text{and, } \Omega = \sqrt{(E+i\Gamma)^2 - \Delta^2} \text{ and for } s\text{- wave } \Delta = \Delta_0. \tag{3}$$

For *d*-wave and *p*-wave, the transmitted hole-like quasiparticle and electron-like quasiparticle experience different effective pair potentials $\Delta(\theta_+)$ and $\Delta(\theta_-)$, respectively shown in Fig. 2(a), with $\theta_+ = \theta$ and $\theta_- = \pi - \theta$. Here, the $\widehat{k_x}$ axis is defined to be the direction along which the magnitude of the pair potential becomes maximum, and the crystal $\hat{n}$ axis is taken along the $\widehat{k_x}$ axis when the symmetry is $d_{x^2-y^2}$ wave. In this situation to capture the essential physics, we have to consider a 2-dimensional model. The effective pair potentials $\Delta(\theta_+)$ and $\Delta(\theta_-)$, for the $d_{x^2-y^2}$ -wave case, can be expressed as $\Delta(\theta_+) = \widetilde{\Delta} Cos(2\theta - 2\alpha)$ and $\Delta(\theta_-) = \widetilde{\Delta} Cos(2\theta + 2\alpha)$ [55]. For *p*-wave symmetry, we took Anderson-Brinkman-Morel (ABM) model [28,58], where the transmitted hole-like quasiparticle and electron-like quasiparticle effective pair potentials can be expressed as $\Delta(\theta_+) = \widetilde{\Delta} Sin(\theta - \alpha)$ and $\Delta(\theta_-) = \widetilde{\Delta} Sin(\theta + \alpha)$, respectively. For the modified BTK model for *p* and *d*-wave proposed by Tanaka, the $a(E, \Delta, \Gamma, Z)$ and $b(E, \Delta, \Gamma, Z)$ are replaced by modified $a(E, \widetilde{\Delta}, \Gamma, Z, ,\theta, \alpha)$ and $b(E, \widetilde{\Delta}, \Gamma, Z, ,\theta, \alpha)$ shown in Eqn. (4), taken from ref. [55];

$$a(E, \widetilde{\Delta}, \Gamma, \theta, Z, \alpha)$$
$$= \frac{4\cos^2\theta \sqrt{(E+i\Gamma) + \Omega_-}\sqrt{(E+i\Gamma) - \Omega_+}e^{-i\varphi_+}}{(4\cos^2\theta + Z^2)\sqrt{(E+i\Gamma) + \Omega_-}\sqrt{(E+i\Gamma) + \Omega_+} - Z^2\sqrt{(E+i\Gamma) - \Omega_-}\sqrt{(E+i\Gamma) - \Omega_+}e^{i(\varphi_- - \varphi_+)}}$$

$$b(E, \widetilde{\Delta}, \Gamma, \theta, Z, \alpha)$$
$$= \frac{-Z(Z + 2i\cos\theta)\left\{\sqrt{(E+i\Gamma) + \Omega_-}\sqrt{(E+i\Gamma) + \Omega_+} - \sqrt{(E+i\Gamma) - \Omega_-}\sqrt{(E+i\Gamma) - \Omega_+}e^{i(\varphi_- - \varphi_+)}\right\}}{(4\cos^2\theta + Z^2)\sqrt{(E+i\Gamma) + \Omega_-}\sqrt{(E+i\Gamma) + \Omega_+} - Z^2\sqrt{(E+i\Gamma) - \Omega_-}\sqrt{(E+i\Gamma) - \Omega_+}e^{i(\varphi_- - \varphi_+)}}$$

$$\text{where, } \Omega_\pm = \sqrt{(E+i\Gamma)^2 - |\Delta(\theta_\pm, \alpha)|^2} \text{ and } e^{i\varphi_\pm} = \Delta(\theta_\pm, \alpha)/|\Delta(\theta_\pm, \alpha)| \quad (4)$$

**Appendix 2**

Here Fig 7. (a) to (f) show all the spectra of the Fig 3 (a) along with their corresponding fitted curves for both *p* wave (in green line) and *d* wave (in blue line). Fig. 7 (g-h) show the fitted value of $\widetilde{\Delta}$, $\Gamma$, Z and α as a function of contact resistance $R_C$ extracted from the *p* wave and *d* wave fits. We observe that $\widetilde{\Delta}$ and α are independent of $R_C$ confirming the reliability of these values. The difference in the height of central conductance peak arises from differences in $\Gamma$ and Z [53,56]; while Z shows a marginal decrease with $R_C$, $\Gamma$ shows a systematic increase. The systematic increase in $\Gamma$ could have from multiple origins. In principle, from quantum mechanical uncertainty principle when the electron passes through a point contact of diameter *D*, the transverse momentum (parallel to the plane of contact) will have an uncertainty $\Delta p = \hbar \Delta k \sim h/D$. This would correspond to an energy uncertainty of the order of $\hbar^2 \Delta k^2/2m$ which will reflect in a corresponding broadening of the spectra. However, a rough quantitative estimate suggests that this broadening should be pronounced only below 100 mK for the contact sizes under consideration, and therefore, cannot be the dominant contribution in our experiment. Another origin of Γ could be from the irregularity in the edges of the point contact, which in practice is assumed to be a neat circle. This irregularity will have more pronounced effect as the point contact diameter is reduced and could contribute to larger values of Γ. Nevertheless, the precise origin for the increase in Γ will have to be established from future theoretical investigations.

**Acknowledgement**

We are thankful to Suman Nandi for helping us with the Lau's diffraction measurement. We would like to thank Department of Atomic Energy, Government of India, for the financial support. We are also thankful to Prof. Yogesh Singh (IISER Mohali) for permitting us to utilize the facilities for the single-crystal growth. A.V. acknowledges the Department of Science and Technology (DST), Government of India, for the Inspire Faculty Fellowship (IFA21-PH279).

**Data Availability**

The data used generated in this work are available from the authors upon request.

[40]M. Smidman, F. K. K. Kirschner, D. T. Adroja, A. D. Hillier, F. Lang, Z. C. Wang, G. H. Cao and S. J. Blundell, *Nodal multigap superconductivity in $KCa_2Fe_4As_4F_2$*, Phys. Rev. B, 97(6), 060509 (2018).

[41]H. Fukazawa, Y. Yamada, K. Kondo, T. Saito, Y. Kohori, K. Kuga, Y. Matsumoto, S. Nakatsuji, H. Kito, P. M. Shirage, K. Kihou, N. Takeshita, C. H. Lee, A. Iyo, and H. Eisaki, *Possible multiple gap superconductivity with line nodes in heavily hole-doped superconductor $KFe_2As_2$ studied by $^{75}As$ nuclear quadrupole resonance and specific heat*, J. Phys. Soc. Jpn., 78(8) (2009).

[42]A. Kamlapure, M. Mondal, M. Chand, A. Mishra, J. Jesudasan, V. Bagwe, L. Benfatto, V. Tripathi, P. Raychaudhuri, *Measurement of magnetic penetration depth and superconducting energy gap in very thin epitaxial NbN films,* Appl. Phys. Lett. 96, 072509 (2010).

[43]P. Das, S. Dutta, Krishna K. S., J. Jesudasan, P. Raychaudhuri, *A compact inertial nanopositioner operating at cryogenic temperatures*, Rev. Sci. Instrum. 95, 113708 (2024).

[44]A. P. Sakhya, A. Pramanik, R. P. Pandeya, S. Datta, A. Singh, A. Thamizhavel, K. Maiti, *Preparation, characterization and X-ray photoemission spectroscopy study of a correlated semimetal, SmBi,* AIP Conf. Proc. 2265, 030369 (2020).

[45]G. T. Woods, R. J. Soulen, Jr., I. Mazin, B. Nadgorny, M. S. Osofsky, J. Sanders, H. Srikanth, W. F. Egelhoff and R. Datla, *Analysis of point-contact Andreev reflection spectra in spin polarization measurements,* Phys. Rev. B 70, 054416 (2004).

[46]G. Sheet, S. Mukhopadhyay, S. Soman and P. Raychaudhuri. *Anomalous structures in point contact Andreev reflection spectrum*, Physica B 359–361, 491 (2005).

[47]G. Sheet, S. Mukhopadhyay, and P. Raychaudhuri, *Role of critical current on the point-contact Andreev reflection spectra between a normal metal and a superconductor*, Phys. Rev. B 69, 134507 (2004).

[48]R. Kumar and G. Sheet, *Nonballistic transport characteristics of superconducting point contacts*, Phys. Rev. B 104, 094525 (2021).

[49]Y. V. Sharvin, *A possible method for studying fermi surfaces*, J. Exptl. Theoret. Phys. (U.S.S.R.) 48, 984-985 (1965).

[50]G. Wexler, *The size effect and the non-local Boltzmann transport equation in orifice and disk geometry*, Proc. Phys. Soc. London 89, 927 (1966).

[51]M. Mondal, B. Joshi, S. Kumar, A. Kamlapure, S. C. Ganguli, A. Thamizhavel, S. S. Mandal, S. Ramakrishnan and P. Raychaudhuri, *Andreev bound state and multiple energy gaps in the noncentrosymmetric superconductor BiPd*, Phys. Rev. B 86, 094520 (2012).

[52]G. E. Blonder and M. Tinkham, *Metallic to tunneling transition in Cu-Nb point contacts*, Phys. Rev. B 27, 112 (1983).

[53]G. Sheet, "*Point contact Andreev reflection spectroscopy on superconductors and ferromagnets*," Ph.D thesis, Tata Institute of Fundamental Research, 2006, available at http://hdl.handle.net/10603/409007.

[54]R. C. Dynes, V. Narayanamurti and J. P. Garno, *Direct measurement of quasiparticle-lifetime broadening in a strong-coupled superconductor,* Phys. Rev. Lett. 41, 1509 (1978).

**Table 1. Fitting parameters for *s*, *p* and *d* wave BTK-Tanaka model at the lowest temperatures for various contacts.**

| Data no. | T (K) | $R_C$ (Ω) | Type of model | $\Delta_0$ or $\tilde{\Delta}$ (mV) | Γ (mV) | α (Degree) | Z |
|---|---|---|---|---|---|---|---|
| 1 | 2.48 | 0.28 | *s* wave | 0.58 | 0.14 | - | 0.93 |
| | | | *p* wave ($p_y$) | 1.00 | 0.001 | 13.00° | 0.88 |
| | | | *d* wave ($d_{x^2-y^2}$) | 0.82 | 0.030 | 12.56° | 1.35 |
| 2 | 2.50 | 0.19 | *s* wave | 0.10 | 0.025 | - | 0 |
| | | | *p* wave ($p_y$) | 0.94 | 0.130 | 29.89° | 2.77 |
| | | | *d* wave ($d_{x^2-y^2}$) | 0.95 | 0.150 | 23.50° | 3.10 |
| 3 | 2.50 | 0.17 | *s* wave | 0.10 | 0.025 | - | 0 |
| | | | *p* wave ($p_y$) | 0.94 | 0.130 | 29.89° | 2.77 |
| | | | *d* wave ($d_{x^2-y^2}$) | 0.99 | 0.155 | 23.50° | 3.20 |
| 4 | 2.50 | 0.24 | *s* wave | 0.10 | 0.030 | - | 0 |
| | | | *p* wave ($p_y$) | 1.00 | 0.250 | 31.59° | 3.40 |
| | | | *d* wave ($d_{x^2-y^2}$) | 1.00 | 0.250 | 24.50° | 3.60 |
| 5 | 2.50 | 0.34 | *s* wave | 0.08 | 0.005 | - | 0 |
| | | | *p* wave ($p_y$) | 0.96 | 0.100 | 30.40° | 4.50 |
| | | | *d* wave ($d_{x^2-y^2}$) | 0.95 | 0.100 | 24.50° | 6.20 |
| 6 | 2.06 | 0.18 | *s* wave | 0.15 | 0.005 | - | 0 |
| | | | *p* wave ($p_y$) | 1.00 | 0.080 | 34.50° | 4.00 |
| | | | *d* wave ($d_{x^2-y^2}$) | 0.99 | 0.060 | 26.50° | 4.00 |

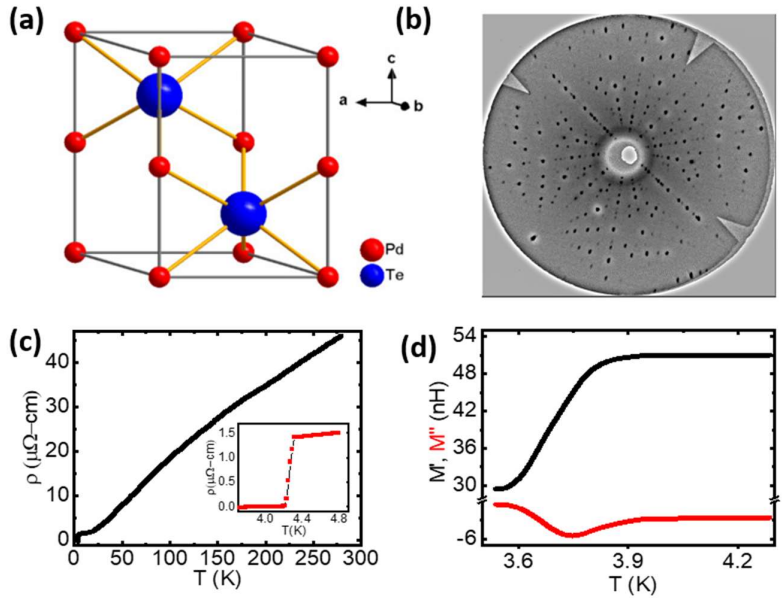

**Fig. 1**. (a) The crystal structure of PdTe, where Pd atoms are shown as red spheres and Te atoms as blue spheres. This image is taken from reference [31]. (b) Laue diffraction pattern of the PdTe crystal. (c) Electrical resistivity as a function of temperature from 2 to 300 K. Inset: The low-temperature electrical resistivity of PdTe exhibits zero resistance below 4.2 K. (d) Real and Imaginary part of mutual inductance (M' and M'') as a function of temperature for a PdTe crystal with superconducting transition, $T_c \sim 3.9$ K.

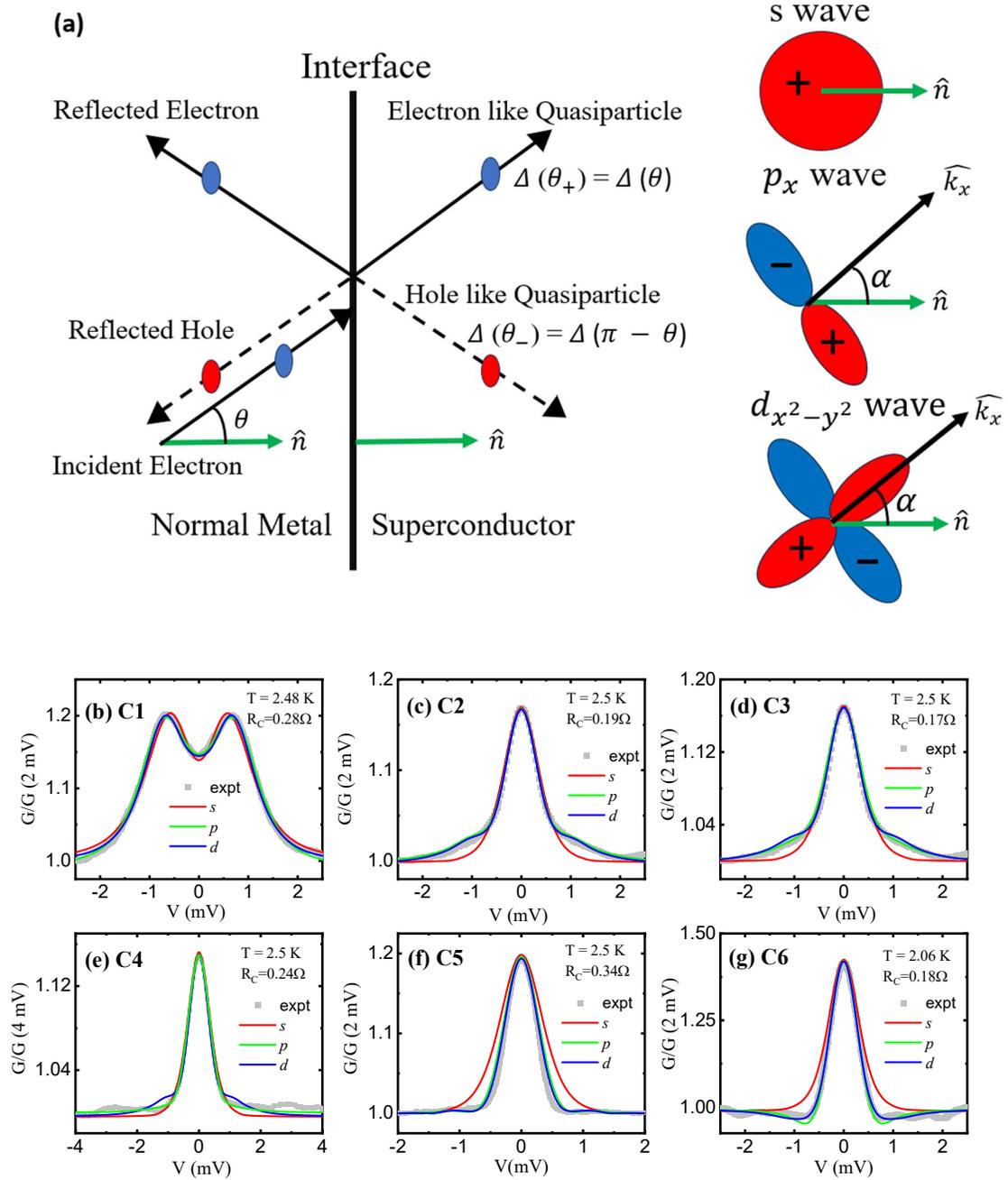

**Fig. 2.** (a) Schematic representation of reflection and transmission phenomena occurring at the interface. For an interface with finite barrier parameter, Z, the incident electron from the normal metal gets partially Andreev (retro-) reflected as a hole and partially reflected as an electron, whereas electron and hole like quasiparticles propagate inside the superconductor. Here $\theta$ is the angle of incidence of the electron with respect to the normal at the interface, $\hat{n}$ and $\alpha$ is the angle between $\widehat{k_x}$ and $\hat{n}$ in the superconductor. (b) to (g) Six different kinds of low temperature PCAR spectra obtained for different contacts established between the sample and the tip. The solid lines show the fit of the spectra with $s$, $p$ and $d$ wave symmetries.

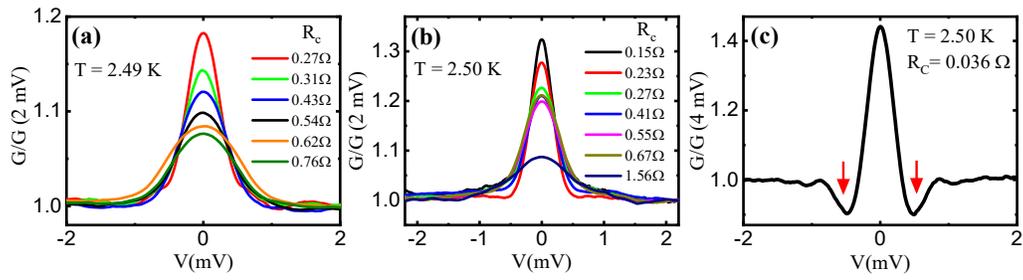

**Fig. 3** (a) - (b) Spectra obtained by changing contact resistance by gradually withdrawing the tip from the sample in small steps. The height of the zero bias conductance peak decreases with increasing contact resistance. (c) Normalized conductance spectra showing critical current feature for very low contact resistance (shown by red arrow).

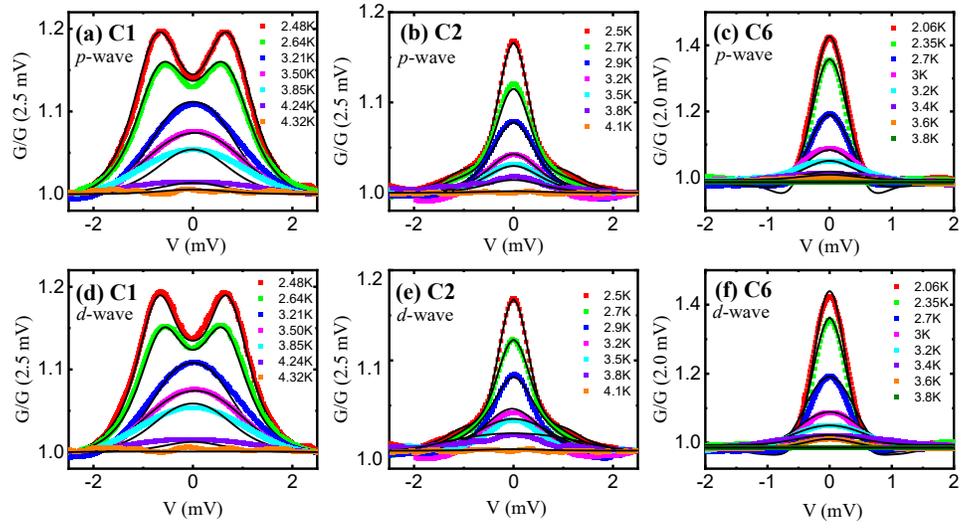

**Fig. 4.** Temperature variation in the PCAR spectra for C1, C2, C6 and fitted with (a)-(c) *p*-wave and, (d)-(f) *d*-wave symmetries.

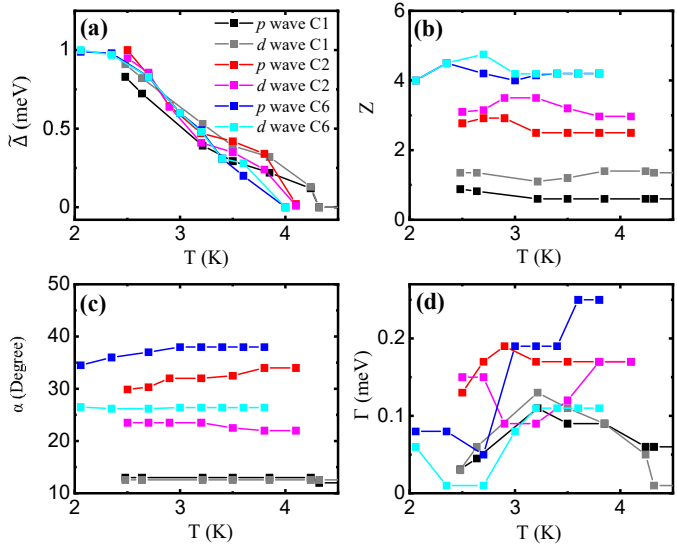

**Fig. 5.** The temperature dependence of $\widetilde{\Delta}$, $\Gamma$, $\alpha$ and $Z$ extracted from the fits shown in Fig 4 assuming *p*-wave and *d*-wave symmetries.

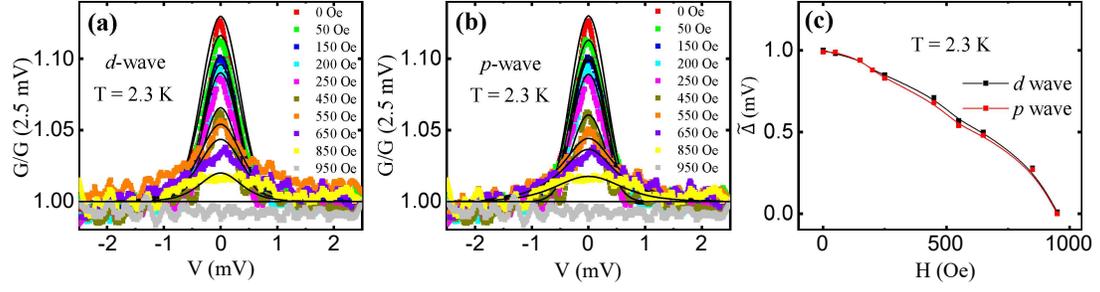

**Fig. 6.** Magnetic-field dependence of PCAR spectra at 2.3 K with solid lines showing (a) *p*-wave fits and (b) *d*- wave fits. (c) Variation of $\widetilde{\Delta}$ with magnetic field obtained from the two fits.

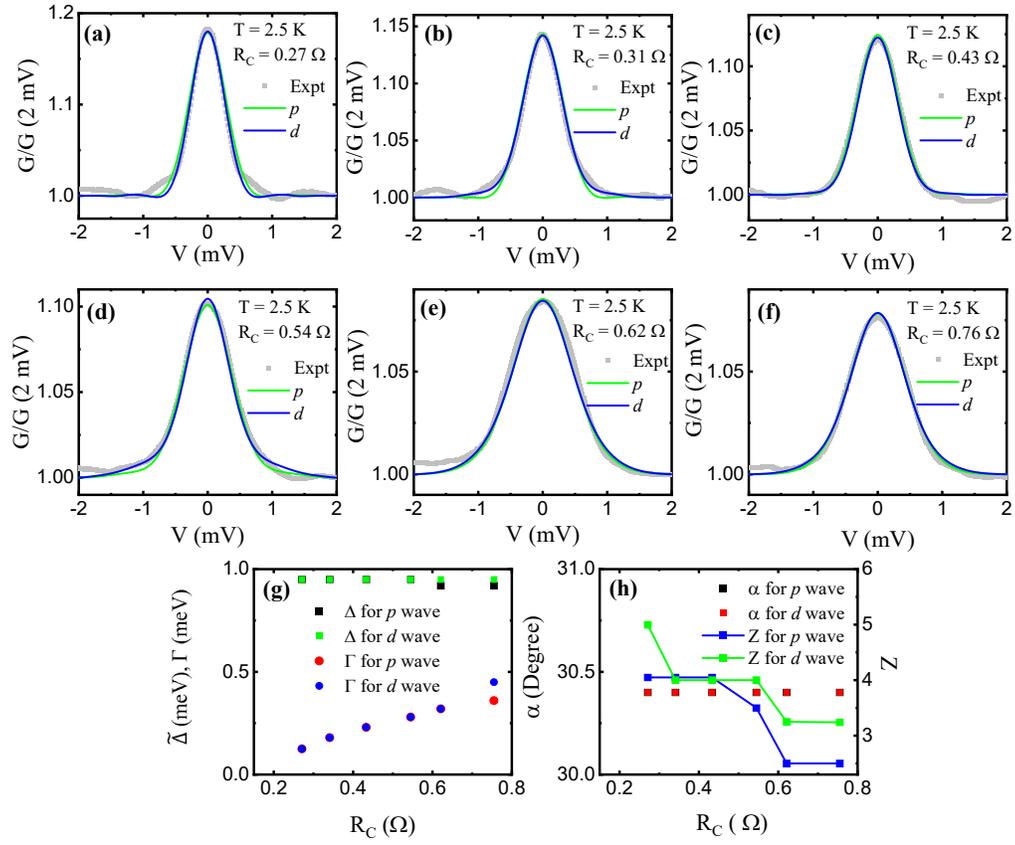

**Fig. 7.** (a)-(f) shows the *p* and *d* wave fits for the set of PCAR spectra shown in Fig. 3 (a) individually. (g) $\widetilde{\Delta}$, $\Gamma$ and (h) Z, α as a function of $R_C$ extracted from the fits.